\documentclass[manuscript]{acmart}

\AtBeginDocument{%
  \providecommand\BibTeX{{%
    \normalfont B\kern-0.5em{\scshape i\kern-0.25em b}\kern-0.8em\TeX}}}

\setcopyright{acmcopyright}
\copyrightyear{2018}
\acmYear{2018}
\acmDOI{XXXXXXX.XXXXXXX}

\acmJournal{JACM}
\acmVolume{37}
\acmNumber{4}
\acmArticle{111}
\acmMonth{8}

\acmPrice{15.00}
\acmISBN{978-1-4503-XXXX-X/18/06}

\usepackage{multirow}
\usepackage[normalem]{ulem}
\useunder{\uline}{\ul}{}
\begin{document}

\title{Improving Sequential Recommendation Models with an Enhanced Loss Function}

\author{Fangyu Li}
\email{lifangyu@stu.xmu.edu.cn}
\author{Shenbao Yu}
\email{yushenba@stu.xmu.edu.cn}
\author{Feng Zeng}
\email{zengfeng@xmu.edu.cn}
\author{Fang Yang}
\authornote{Corresponding author}
\email{yang@xmu.edu.cn}
\affiliation{%
  \institution{Department of Automation, Xiamen University}
  \city{Xiamen}
  \country{China}
}




\renewcommand{\shortauthors}{Trovato and Tobin, et al.}

\begin{abstract}
There has been a growing interest in benchmarking sequential recommendation models and reproducing/improving existing models. For example, Rendle et al. improved matrix factorization models by tuning their parameters and hyperparameters. Petrov and Macdonald developed a more efficient and effective implementation of BERT4Rec, which resolved inconsistencies in performance comparison between BERT4Rec and SASRec in previous works.
In particular, BERT4Rec and SASRec share a similar network structure, with the main difference lying in their training objective/loss function. 
Therefore, we analyzed the advantages and disadvantages of commonly used loss functions in sequential recommendation and proposed an improved loss function that leverages their strengths.
We conduct extensive experiments on two influential open-source libraries, and the results demonstrate that our improved loss function significantly enhances the performance of GRU4Rec, SASRec, SR-GNN, and S3Rec models, improving their benchmarks significantly. Furthermore, the improved SASRec benchmark outperforms BERT4Rec on the ML-1M and Beauty datasets and achieves similar results to BERT4Rec on the ML-20M and Steam datasets. We also reproduce the results of the BERT4Rec model on the Beauty dataset. Finally, we provide a comprehensive explanation of the effectiveness of our improved loss function through experiments.
Our code is publicly available at \url{https://github.com/Li-fAngyU/sequential_rec}.
\end{abstract}



\maketitle

\section{Introduction}
With the development of deep learning, a plethora of neural network-based models have been proposed in the field of sequential recommendation. Notably, models based on the Transformer network architecture have gained the most popularity, including SASRec \cite{ICDM-2018-kang}, BERT4Rec \cite{BERT4Rec}, S$^3$-Rec \cite{S3Rec}, and NOVA-BERT \cite{NOVABERT}.
However, variations in data preprocessing, evaluation metrics, and model implementation details have led to inconsistent performance of the same model across different research studies.
Therefore, in recent years, there has been a growing interest in the benchmark of sequential recommendation, including efforts to improve evaluation metrics, optimize dataset partitioning strategies, and facilitate model results replication \cite{SysRec-2021-Dallmann, RSS, KDD-2020-Rendle, RecSys20-Data-Splitting-Strategies, RecSys22-iALS}.
Regarding model replication, Rendle et al. \cite{RecSys22-iALS} demonstrated the effectiveness of the iALS loss function and embedding dimension adjustments in improving the performance of traditional matrix factorization models.
In a systematic analysis and replication study of BERT4Rec \cite{2022BERT4Rec}, Petrov and Macdonald presented a more effective and efficient implementation than that described in the original paper.

Drawing inspiration from these works, we have initiated an investigation into potential optimizations of classical sequential recommendation models, such as GRU4Rec \cite{ICLR-2015-hidasi}, SR-GNN \cite{SRGNN} and SASRec \cite{ICDM-2018-kang}. 
Intuitively, we attempt to initiate our approach from the perspective of the training objective or loss function. 
This is motivated by several reasons. 
Firstly, there are numerous loss functions available for sequential recommendation, such as Cross-Entropy (CE), Binary Cross-Entropy (BCE), Bayesian Personalized Ranking (BPR) \cite{UAI-2009-Rendle}, TOP1 \cite{ICLR-2015-hidasi}, Masked Language Model loss (MLM), etc. 
However, prior research has primarily focused on ablation experiments involving model hyperparameters, with relatively less attention given to the selection and experimental analysis of loss functions. 
Secondly, while SASRec and BERT4Rec have similar model architectures, their respective loss functions serve as the primary distinguishing factor between the two. Additionally, some studies have indicated that BERT4Rec's use of an item masking \cite{BERT} training task provides a significant advantage over SASRec \cite{BERT4Rec, RSS}.
Thirdly, the performance of the model is significantly impacted by the training target or loss function, even without altering the model architecture.

\begin{figure}[t]
  \centering
  \includegraphics[width=0.7\linewidth]{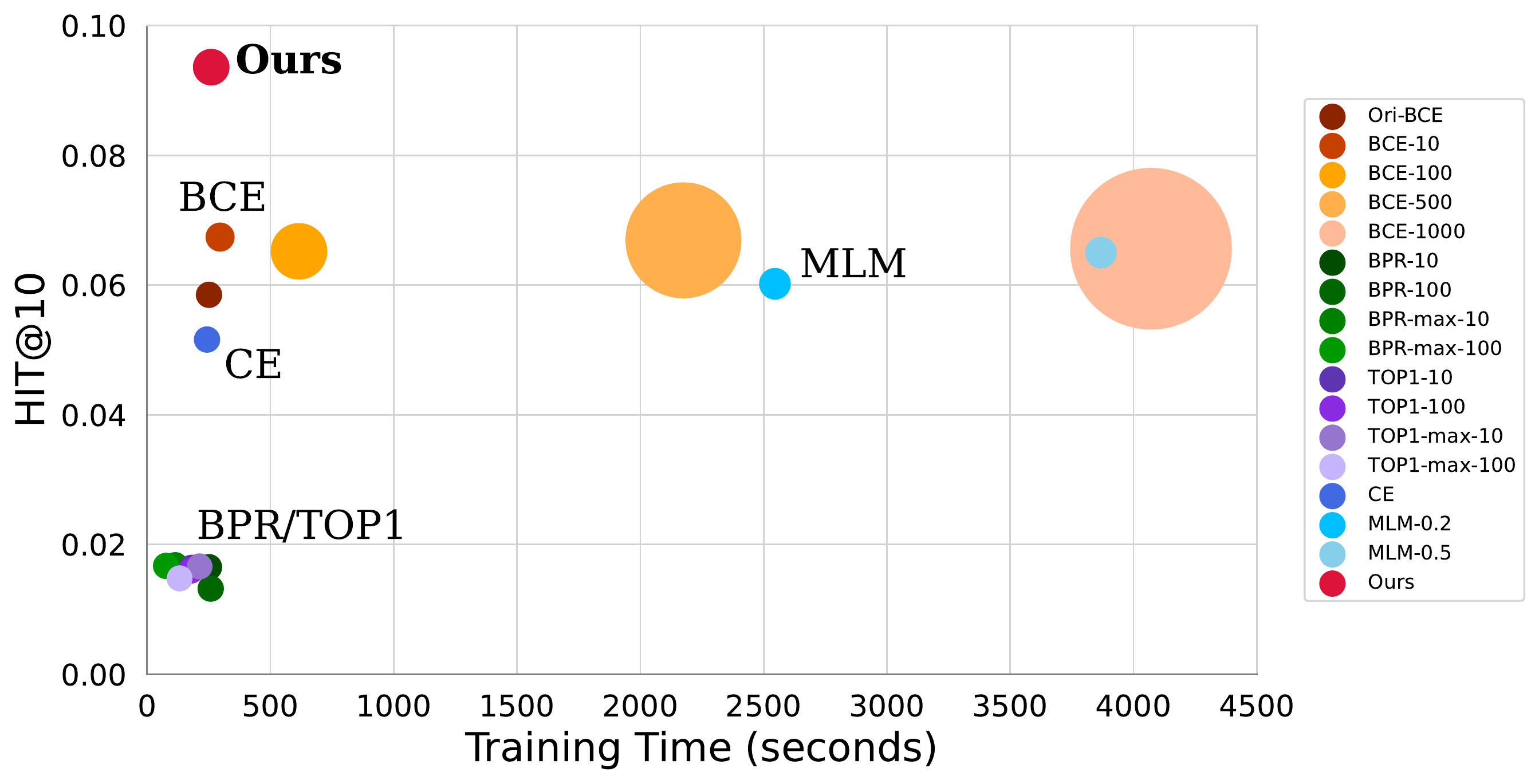}
  \caption{Comparison of SASRec's performance on the Beauty dataset using different loss functions, including BCE, CE, BPR, BPR-max, TOP1, TOP1-max, MLM, and our enhanced loss function. Our enhanced loss function outperforms the others in terms of full rank metric HIT@10, with shorter training time and lower GPU memory consumption. The size of each bubble corresponds to the magnitude of GPU memory consumption during the training process. The numbers following BCE, BPR, BPR-max, TOP1, and TOP1-max denote the number of negative instances used in training.}
  \label{fig:introduction}
\end{figure}

Therefore, We analyzed the advantages and disadvantages of
commonly used loss functions in sequential recommendation and proposed an improved loss function that leverages their strengths.
Figure \ref{fig:introduction} briefly summarises the results of our findings on the Beauty dataset.
The figure illustrates the performance comparison of SASRec using different loss functions, where the observed HIT@10 effectiveness varies from 0.0168 (BPR-max-10) to 0.0936 (our improved loss function), demonstrating a 5.5-fold difference.
The figure also indicates that BCE-10, BCE-500, etc., can outperform Ori-BCE (original BCE used in SASRec with only one negative sample).
However, due to the fact that BCE involve all timestamp of the train sequence, increasing the number of negative samples leads to a substantial rise in training time (258s for BPR-100, 616s for BCE-100, 4071s for BCE-1000) and memory consumption (285MB for BPR-100, 1507MB for BCE-100, 12893MB for BCE-1000) in comparison to BPR.

Ultimately, We conduct extensive experiments on two influential open-source libraries. The experimental results demonstrate that our improved loss function significantly improves the performance of various network structures, including GRU4Rec, SASRec, S$^3$-Rec, and SR-GNN.
Futhermore, when trained with our enhanced loss function, SASRec outperforms BERT4Rec in ML-1M and Beauty datasets.
Petrov and MacDonald were unable to reproduce the results of BERT4Rec on the Beauty dataset in their work. However, we were able to successfully replicate the results of BERT4Rec on the Beauty dataset by expanding the dataset while using their open-source code for experimentation.
Finally, we provide a comprehensive explanation of the effectiveness of our improved loss function.




\section{Background}
Sequential recommendation is an important technique in recommendation systems that aims to predict the next item or content that a user is likely to be interested in.
Suppose that there are a set of users $ \mathcal{U} = \left( u_1, u_2, ..., u_{ \left| \mathcal{U} \right| }\right) $ and a set of items $ \mathcal{I} = \left( i_1, i_2, ..., i_{ \left| \mathcal{I} \right| }\right) $, where $ \left| \mathcal{U} \right| $ and $ \left| \mathcal{I} \right| $ denote the the number of users and items, respectively.
In the sequential recommendation, we mainly focus on the user's historical interaction records.
Therefore, we formulate a user sequence $S_{1:l} = \left( S_1,S_2, ...,S_{l} \right)$ based on interaction records in chronological order, where $l$ denotes the length of user sequence and $S_t$ denotes the user interaction item at timestamp $t$.
Finally, we can describe a sequential recommendation model $f$ as follow: $R_{l} = f(S_{1:l})$ where $R_{l}=\left(r_{l,1},r_{l,2}, ...,r_{l,\left| \mathcal{I} \right|} \right)$ denotes the outputs of all items at timestamp $l$, where $ r_{l,k} $ is the prediction score of item $i_k$ at timestamp $l$.


The original publication on GRU4Rec presented the model's performance results utilizing BPR, TOP1, and CE loss functions. However, further research has pointed out that BPR and TOP1 suffer from vanishing gradient problem \cite{CIKM-2018-hidasi}. 
To address this issue, BPR-max and TOP1-max have been introduced, utilizing the softmax score to solve the problem of vanishing gradients.
Indeed, subsequent sequential recommendation models have used various loss functions, such as CE \cite{NARM, STAMP, SRGNN}, BCE \cite{ICDM-2018-kang, Caser, S3Rec, RKSA, ELECRec, CAFE}, MLM \cite{BERT4Rec, NOVABERT, 2022BERT4Rec}, and LambdaRank \cite{RSS, microsoft-2010-burges}.
While most research in this field merely mentions the employed loss function without providing further analysis, there have been studies \cite{CIKM-2018-hidasi, RSS} that compare the performance of different loss functions and analyze their impact on the model's performance.
In the next section, we analyze the advantages and disadvantages of loss functions commonly used in the sequential recommendation and propose an improved loss function that combines their strengths.



\section{Loss Function}
\label{sec:exploration}
In this section, we analyzed the advantages and disadvantages of commonly used loss functions in sequential recommendation and finally proposed an improved loss function by combining their strengths. 
To enhance the readability and coherence of the section, we have excluded the regularization terms of the loss function.

\subsection{BPR \& TOP1}
BPR, TOP1, and their variants have been widely used in sequential recommendation \cite{ICLR-2015-hidasi, CIKM-2018-hidasi, GRU4Recf, HGN, MA-GNN, STEN}. 
The formulation of these two classes of loss functions exhibits similarities, as both involve negative sampling and aim to optimize the model parameters by increasing the score differential between positive and negative samples. 
Since BPR and TOP have similar structures, here we only present the mathematical expressions for BPR and BPR-max below:
\begin{align}
    &L_{bpr} = - \frac{1}{N_s}\sum_{j=1}^{N_s} \log \sigma(r_{l,pos}-r_{l,neg_j}), \\
    &L_{bpr-max} = - \log \sum_{j=1}^{N_s} s_{neg_j} \sigma(r_{l,pos}-r_{l,neg_j}),
\end{align}
where $N_s$ represents the overall number of negative samples, while $r_{l,pos}$ and $r_{l,neg}$ refer to the scores assigned to the positive and negative items, respectively, at the final timestamp $l$ of the input sequence. Moreover, $s_{neg}$ indicates the softmax score allocated to the negative examples. Additionally, $\sigma$ represents the sigmoid function.

Figure 2a shows an example that corresponds to BPR and TOP1, where the calculation of the loss function solely involves the last timestamp of the input sequence.
One of the advantages of this approach is that it remains independent of the input sequence's length. Moreover, the GPU memory consumption does not rise considerably when the number of negative samples is increased. 
For instance, as presented in Figure 1, while BCE-100 requires 1507MB of GPU memory usage, BPR-100 necessitates only 285MB. 
However, this approach has its own drawbacks. Specifically, when the item set is extensive, informative negative samples with high scores may go unnoticed due to negative sampling issues.
The issue of gradient vanishing in the original BPR and TOP1 has been extensively described in the research conducted by Hidasi and Karatzoglou \cite{CIKM-2018-hidasi}.

\begin{figure}[t]
  \centering
  \includegraphics[width=0.9\linewidth]{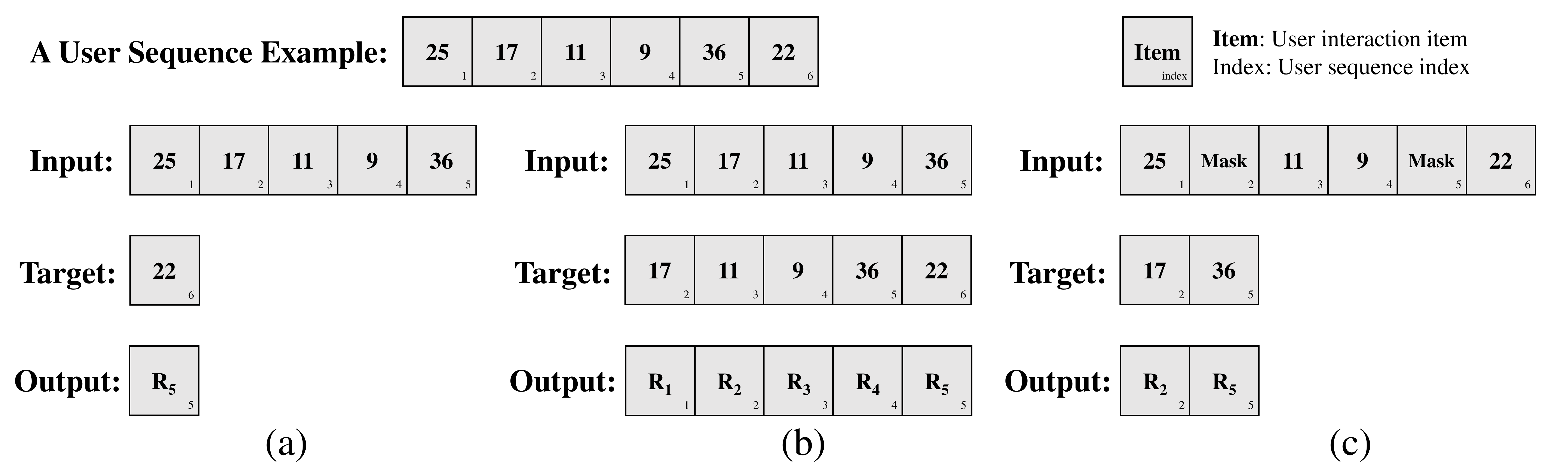}
  \caption{Illustration of the three approaches to optimizing a model using loss functions. Input, Output, and Target respectively denote the model's input, output, and labels.Using a virtual user sequence example, we have demonstrated three approaches to the required model input, output, and labels for loss functions. Figure (a) displays the approach that involves only the last timestamp of the input sequence, while Figure (b) demonstrates the approach that involves all timestamps of the input sequence. Figure (c) exemplifies the MLM used in BERT4Rec. Further analysis and description can be found in Section \ref{sec:exploration}.}
  \label{fig:loss}
\end{figure}

\subsection{BCE}
The binary cross-entropy (BCE) loss function is also a widely used loss function in sequential recommendation \cite{ICDM-2018-kang, Caser, S3Rec, RKSA, ELECRec, CAFE}. Its primary objective is to drive the score of positive samples towards infinity and the score of negative samples towards negative infinity. The specific mathematical formula for BCE is as follows:
\begin{equation}\label{equ:loss_sasrec}
    L_{bce} = -\sum_{t=1}^{l} \left[ \log\sigma(r_{t,pos}) + \sum_{j = 1}^{N_s}\log\sigma(1-r_{t,neg_j}) \right],
\end{equation}
where $l$ denotes the length of the input sequence.
It can be observed that the calculation of the BCE loss involves all timestamps of the input sequence, and the example in Figure 2b can serve as a graphical representation of BCE.
Comparison of Figures 2a and 2b reveals that the primary difference between the BCE and BPR/TOP1 losses is that the optimization objective of BCE involves the entire input sequence. Based on the results shown in Figure 1, where BCE outperforms BPR and TOP1, we hypothesize that the optimization objective that incorporates the entire input sequence is the primary reason for this performance improvement. Therefore, it can be concluded that this approach is the most significant advantage of BCE.
The issue of negative sampling is also present in BCE and is even more difficult to handle. 
In BCE, negative sampling must be performed at each timestamp of the sequence, which results in a larger sampling quantity compared to BPR. 
As a result, BCE requires more training time and GPU memory. 
For instance, if the sequence length is 50, BPR samples 50 negative examples, while BCE requires the sampling of 2,500 negative examples.
From Figure 1, it can be observed that as the number of negative samples in BCE increases, the GPU memory used by the model significantly increases.


\subsection{CE \& MLM}
The cross-entropy (CE) loss function is a popular choice in sequence recommendation due to its computational simplicity, fast convergence, and ability to address the issue of vanishing gradients while taking into account class relationships. Prior studies in this field have extensively utilized the CE loss function \cite{NARM,STAMP, ICLR-2015-hidasi, SRGNN}, employing it in the manner depicted in Figure \ref{fig:loss}a.

The masked language modeling (MLM) loss, which includes the use of CE, has gained significant attention in the area of sequence recommendation in recent years \cite{BERT4Rec, NOVABERT, 2022BERT4Rec}. Figure \ref{fig:loss}c illustrates the graphical representation of MLM, and its primary training objective is to predict the current interaction item based on the user's past and future interaction information. Although MLM has shown remarkable performance in other fields \cite{BERT, vit, swin}, its effectiveness in the context of sequential recommendation may be limited due to the inability to know the user's future interaction items in actual scenarios. Nonetheless, recent work such as BERT4Rec has leveraged MLM for training and achieved state-of-the-art performance in sequential recommendation \cite{2022BERT4Rec}. 
We believe that one of the salient features of BERT4Rec that we can learn from is its application of CE to any timestamp of the input sequence at random.

\subsection{Our Enhanced Loss Function}
Finally, we propose an enhanced loss function that addresses the limitations of applying CE solely to the last timestamp of the input sequence and leverages the advantages of optimizing all timestamps in BCE.


\begin{equation}
    L = - \frac{1}{l} \sum_{t=1}^{l} \log \frac{exp\left( r_{t,pos} \right)}{\sum_{ j=1 }^{\left| \mathcal{I} \right|}exp\left( r_{t,j} \right)}.
\end{equation}

This improved loss function is simple and straightforward, as reflected in the formula, which adds a cumulative term for time before the CE loss. Since the model structure and loss function are interrelated, some sequential recommendation models fuse the information of the input sequence and only output the prediction for the next item, such as NARM \cite{NARM}, STAMP \cite{STAMP}, SR-GNN \cite{SRGNN}, etc. These models cannot directly use the improved loss function without modifications to the model structure. In Section \ref{sec:experiments}, we conduct experiments by making simple modifications to the network structure of SR-GNN to evaluate the efficacy of the proposed enhanced loss function.

In our initial research on loss function, we experimented with increasing the number of negative samples in the BCE loss function, which yielded slight improvements. However, as the number of negative samples increased, the model's performance plateaued, while requiring more training time and GPU memory. Subsequently, we explored other loss functions commonly used in sequential recommendation and found that CE could achieve similar performance to BCE (as illustrated in Figure \ref{fig:introduction}). We then extended CE to all timestamps of the input sequence, which resulted in significant improvements to the SASRec model. Notably, when we applied the proposed enhanced loss function to other models, including GRU4Rec, S$^3$Rec, and SR-GNN, we observed significant performance gains. For a detailed analysis of our experimental results, please refer to the next section.


\section{Experiments}
\label{sec:experiments}

Figure \ref{fig:introduction} shows the preliminary experimental results of the SASRec model on the Beauty dataset, which demonstrates the significant improvement brought by our improved loss function. To further validate and explain the effectiveness of our proposed approach, we conducted experiments to answer the following two research questions:
\begin{itemize}
    \item RQ1: Can our improved loss function lead to better recommendation results on different datasets and models?
    \item RQ2: What is the reason for the significant performance improvement achieved by our proposed loss function?
\end{itemize}

\subsection{Experiment Setup}
Modifying the loss function directly on the source code of the model ensures the accuracy of the experimental results. For our experiments, we selected the source code provided by S$^3$Rec (S3Rec repo\footnote{ \url{https://github.com/aHuiWang/CIKM2020-S3Rec}}) \cite{S3Rec} and Petrov and Macdonald (Aprec repo\footnote{ \url{https://github.com/asash/bert4rec_repro}}) \cite{2022BERT4Rec}. The former can perfectly reproduce the results of the S$^3$Rec and SASRec models as reported in the original paper, while the latter can reproduce the results of the BERT4Rec model on three datasets. Conducting experiments on different code repositories can improve the accuracy and persuasiveness of the experimental results.
Throughout all experiments, the model parameters and training configurations were maintained at their default values.

For dataset selection, we used the datasets used in the original code to avoid bias caused by inconsistent dataset preprocessing. The S3Rec repository (S3Rec repo) contains five datasets, namely Yelp, LastFM \cite{LastFM-dataset}, Amazon Beauty$_1$, Sports, and Toys \cite{amazon-datasets}. The Aprec repository (Aprec repo) contains four datasets, namely ML-1M \cite{ml-datasets}, Beauty$_2$ \cite{amazon-datasets}, Steam \cite{ICDM-2018-kang}, and ML-20M \cite{ml-datasets}. Although both code repositories contain the Beauty dataset, the statistical information they provide is different because they use different dataset preprocessing methods. Table \ref{table:statistics of datasets} provides detailed information on each dataset.

\begin{table}[]
\caption{Statistics of datasets after preprocessing}
\resizebox{0.8\textwidth}{!}{
\begin{tabular}{llccccclccccl}
\hline
\multirow{2}{*}{Dataset} &  & \multicolumn{5}{c}{S3Rec repo}             &  & \multicolumn{4}{c}{Aprec repo}       &  \\ \cline{3-7} \cline{9-12}
                         &  & Yelp   & LastFM & Beauty$_1$ & Sports & Toys   &  & ML-1M  & Beauty$_2$ & Steam   & ML-20M   &  \\ \hline
Users                    &  & 30431  & 1090   & 22362  & 35598  & 19412  &  & 6040   & 40226  & 281428  & 138493   &  \\
Items                    &  & 20033  & 3646   & 12101  & 18357  & 11924  &  & 3416   & 54542  & 13044   & 26744    &  \\
Interactions             &  & 316454 & 52551  & 198502 & 296337 & 167597 &  & 999611 & 353962 & 3488885 & 20000263 &  \\
Average Length           &  & 15.80  & 14.41  & 16.40  & 16.14  & 14.06  &  & 165.49 & 8.79   & 12.40   & 144.41   &  \\
Density                  &  & 0.05\% & 1.32\% & 0.07\% & 0.05\% & 0.07\% &  & 4.85\% & 0.02\% & 0.10\%  & 0.54\%   &  \\ \hline
\end{tabular}}
\label{table:statistics of datasets}
\end{table}

For model selection, the S3Rec repo includes implementations of both S$^3$Rec and SASRec models. The former is a pre-trained model based on SASRec, while the latter utilizes the Transformer architecture. To ensure a diverse set of models, we have extended the repository to include GRU4Rec, an RNN-based model, and SR-GNN, a GNN-based model. The implementations of the latter two models are based on the well-known open-source recommendation algorithm development framework, RecBole\footnote{\url{https://github.com/RUCAIBox/RecBole}}. With regards to SR-GNN, we have made minor modifications to its network architecture, enabling it to generate predictions for all timestamps.

The Aprec repo contains implementations of both SASRec and multiple variants of the BERT4Rec models. Nevertheless, the BERT4Rec model implementation in this repository fails to reproduce the experimental results reported in the original BERT4Rec paper on the Beauty dataset. We believe this discrepancy arises from differences in the data preprocessing techniques used. To rectify this issue, we have expanded the Aprec repo to include experiments on the Beauty dataset preprocessed using the S3Rec repo. 


For evaluation metrics, we use two commonly used evaluation metrics for sequential recommendation: top-k Hit Ratio (HIT@$k$) and top-k Normalized Discounted Cumulative Gain (NDCG@$k$). These metrics are widely used in the literature to evaluate the quality of recommendation algorithms for sequential data. Following the recommendations of \cite{KDD-2020-Rendle, SysRec-2021-Dallmann, Recsys-2020-Pablo}, we use full ranking evaluation metrics for our experiments. This entails ranking all the items in the dataset and evaluating the recommendations based on this ranking.

\begin{table}[t]
\caption{Using S3Rec repository, the comparison of full rank metrics HIT@10 and NDCG@10 results on different datasets using original and improved loss functions with four sequential recommendation models. \uline{Underline} indicates the best performance. \textbf{Bold} indicates performance improvement with the improved loss function.}
\resizebox{0.8\textwidth}{!}{
\begin{tabular}{lclcccclcccc}
\hline
\multirow{2}{*}{Dataset} & \multirow{2}{*}{Metrics} &  & \multicolumn{4}{c}{Training using original loss function} &  & \multicolumn{4}{c}{Training using improved loss function}                                                           \\ \cline{4-7} \cline{9-12} 
                         &                          &  & GRU4Rec       & SR-GNN       & SASRec      & S3Rec       &  & GRU4Rec                   & SR-GNN                    & SASRec                   & S3Rec                           \\ \hline
\multirow{2}{*}{Yelp}    & HIT@10                   &  & 0.0164        & 0.0121       & 0.0273      & 0.0354      &  & \textbf{0.0367(123.78\%)} & \textbf{0.0194(60.33\%)}  & \textbf{0.0379(38.83\%)} & {\ul \textbf{0.0474(33.90\%)}}  \\
                         & NDCG@10                  &  & 0.0078        & 0.0061       & 0.0140      & 0.0173      &  & \textbf{0.0184(135.90\%)} & \textbf{0.0099(62.30\%)}  & \textbf{0.0198(41.43\%)} & {\ul \textbf{0.0243(40.46\%)}}  \\ \hline
\multirow{2}{*}{LastFM}  & HIT@10                   &  & 0.0312        & 0.0165       & 0.0404      & 0.0688      &  & \textbf{0.0459(47.12\%)}  & \textbf{0.0394(138.79\%)} & \textbf{0.0587(45.30\%)} & {\ul \textbf{0.0789(14.68\%)}}  \\
                         & NDCG@10                  &  & 0.0171        & 0.0094       & 0.0223      & 0.0356      &  & \textbf{0.0262(53.22\%)}  & \textbf{0.0233(147.87\%)} & \textbf{0.0354(58.74\%)} & {\ul \textbf{0.0381(7.02\%)}}   \\ \hline
\multirow{2}{*}{Beauty}  & HIT@10                   &  & 0.0343        & 0.0278       & 0.0573      & 0.0614      &  & \textbf{0.0695(102.62\%)} & \textbf{0.0652(134.53\%)} & \textbf{0.0932(62.65\%)} & {\ul \textbf{0.1031(67.92\%)}}  \\
                         & NDCG@10                  &  & 0.0185        & 0.0149       & 0.0305      & 0.0307      &  & \textbf{0.0408(120.54\%)} & \textbf{0.0424(184.56\%)} & \textbf{0.0568(86.23\%)} & {\ul \textbf{0.0619(101.63\%)}} \\ \hline
\multirow{2}{*}{Sports}  & HIT@10                   &  & 0.0163        & 0.0128       & 0.0330      & 0.0359      &  & \textbf{0.0357(119.02\%)} & \textbf{0.0336(162.50\%)} & \textbf{0.0541(63.94\%)} & {\ul \textbf{0.0642(78.83\%)}}  \\
                         & NDCG@10                  &  & 0.0085        & 0.0066       & 0.0184      & 0.0182      &  & \textbf{0.0187(120.00\%)} & \textbf{0.0219(231.82\%)} & \textbf{0.0318(72.83\%)} & {\ul \textbf{0.0371(103.85\%)}} \\ \hline
\multirow{2}{*}{Toys}    & HIT@10                   &  & 0.0153        & 0.0163       & 0.0613      & 0.0641      &  & \textbf{0.0597(290.20\%)} & \textbf{0.0681(317.79\%)} & \textbf{0.0989(61.34\%)} & {\ul \textbf{0.1096(70.98\%)}}  \\
                         & NDCG@10                  &  & 0.0083        & 0.0095       & 0.0347      & 0.0335      &  & \textbf{0.0354(326.51\%)} & \textbf{0.0459(383.16\%)} & \textbf{0.0615(77.23\%)} & {\ul \textbf{0.0664(98.21\%)}}  \\ \hline
\end{tabular}}
\label{table: S3Rec repo}
\end{table}

\begin{table}[]
\caption{Comparison of HIT@10 and NDCG@10 results using Aprec repo. 
 \textbf{SASRec} indicates training using improved loss function.
 BERT4Rec$_{hf}$ and BERT4Rec-VAE indicate different implementations mentioned in \cite{2022BERT4Rec}.}
\resizebox{0.8\textwidth}{!}{
\begin{tabular}{lccccclcc}
\hline
\multirow{2}{*}{Dataset}   & \multirow{2}{*}{Metrics} & \multirow{2}{*}{SASRec} & \multirow{2}{*}{BERT4Rec$_{hf}$} & \multirow{2}{*}{BERT4Rec-VAE} & \multirow{2}{*}{\textbf{SASRec}} &  & Reported \cite{2022BERT4Rec}                               & Reported\cite{BERT4Rec} \\ \cline{8-9} 
                           &                          &                         &                               &                               &                                  &  & \multicolumn{1}{l}{BERT4Rec$_{hf}$/BERT4Rec-VAE} & BERT4Rec        \\ \hline
\multirow{2}{*}{ML-1M}   & HIT@10                   & 0.2089                  & 0.2594                      & 0.2512                        & {\ul \textbf{0.2849(36.38\%)}}   &  & 0.2584/0.2394                               & N/A             \\
                         & NDCG@10                  & 0.1119                  & 0.1409                      & 0.1387                        & {\ul \textbf{0.1642(46.74\%)}}   &  & 0.1392/0.1314                               & N/A             \\ \hline
\multirow{2}{*}{ML-20M}  & HIT@10                   & 0.1437                  & {\ul 0.2378}                & 0.1558                        & \textbf{0.2176(51.43\%)}                  &  & 0.2393/0.2886                               & N/A             \\
                         & NDCG@10                  & 0.0716                  & {\ul 0.1310}                & 0.0892                        & \textbf{0.1214(69.55\%)}                  &  & 0.1310/0.1732                               & N/A             \\ \hline
\multirow{2}{*}{Steam}   & HIT@10                   &       0.1216                  &                      0.1211       &       0.1038           &      {\ul \textbf{0.1343(10.44\%)}}            &  & 0.1361/0.1237        & N/A         \\
                         & NDCG@10                  &            0.0632             &                    0.0636         &      0.0541            &          {\ul \textbf{0.0721(14.08\%)}}       &  & 0.0734/0.0526                               & N/A             \\ \hline
\multirow{2}{*}{Beauty$_2$} & HIT@10           & 0.0055                  & 0.0179                      & 0.0330                        &     {\ul \textbf{0.0433(687.27\%)}}              &  & 0.0166/0.0331                               & N/A             \\
                         & NDCG@10                  & 0.0026                  & 0.0087                      & 0.0187                        &       {\ul \textbf{0.0256(884.62\%)}}        &  & 0.0080/0.0188                               & N/A             \\ \hline
\multirow{4}{*}{Beauty$_1$} & HIT@10                   & 0.0316                  & 0.0344                      & 0.0612                        & {\ul \textbf{0.0879(178.16\%)}}  &  & N/A                                         & N/A             \\
                         & NDCG@10                  & 0.0162                  & 0.0167                      & 0.0357                        & {\ul \textbf{0.0526(224.69\%)}}  &  & N/A                                         & N/A             \\
                         & Sample HIT@10            & 0.2468                  & 0.2726                      & 0.3110                        & {\ul \textbf{0.3810(54.38\%)}}   &  & N/A                                         & 0.3025          \\
                         & Sample NDCG@10           & 0.1218                  & 0.1382                      & 0.1902                        & {\ul \textbf{0.2364(94.09\%)}}   &  & N/A                                         & 0.1862          \\ \hline
\end{tabular}}
\label{table: Aprec repo}
\end{table}

\subsection{RQ1: Evaluating the Effectiveness of Improved Loss Function in Sequential Recommendation}
To answer RQ1, we experimented with our improved loss function in two influential open-source libraries for sequential recommendation: S3Rec repo and Aprec repo.
We first implemented our improved loss function in both libraries and trained their models with the original and our improved loss function.

Table \ref{table: S3Rec repo} demonstrates the effectiveness of our improved loss function, which significantly improves the performance of four models on five datasets under the S3rec repository. These four models include Transformer-based, RNN-based, GNN-based, and pre-training models. Specifically, we made minor modifications to the network architecture of SR-GNN, enabling it to output predictions for all timestamps in a sequence.
The performance improvement of the SR-GNN model achieved by our improved loss function is remarkable, ranging from 60.33\% to 383.16\%.

The experimental results of Petrov and Macdonald \cite{2022BERT4Rec} show that BERT4Rec outperforms SASRec. However, their SASRec model was trained using the original BCE loss. Therefore, we expanded the \textbf{SASRec} model and compared it with BERT4Rec in the Aprec repo. The results in Table \ref{table: Aprec repo} show that \textbf{SASRec} outperforms the BERT4Rec model on the ML-1M and Beauty datasets. Additionally, it achieves results comparable to BERT4Rec on the ML-20M and Steam datasets, with less training time. Furthermore, from the table, we can see that \textbf{SASRec} outperforms SASRec on all datasets, particularly on the Beauty2 dataset, where it is 8.84 times better than SASRec, 1.94 times better than BERT4Rec$_{hf}$, and 0.36 times better than BERT4Rec-VAE. 
Finally, we demonstrate that BERT4Rec-VAE can outperform the original implementation \cite{BERT4Rec} on the Beauty1 dataset, proving that the reason Petrov and Macdonald failed to replicate results on the Beauty datasets was due to data preprocessing issues.

In conclusion, our improved loss function is effective in sequential recommendation. Additionally, we provide reproducible code that can replicate the results in Table \ref{table: S3Rec repo} and Table \ref{table: Aprec repo}.


\begin{figure}[t]
  \centering
  \includegraphics[width=0.7\linewidth]{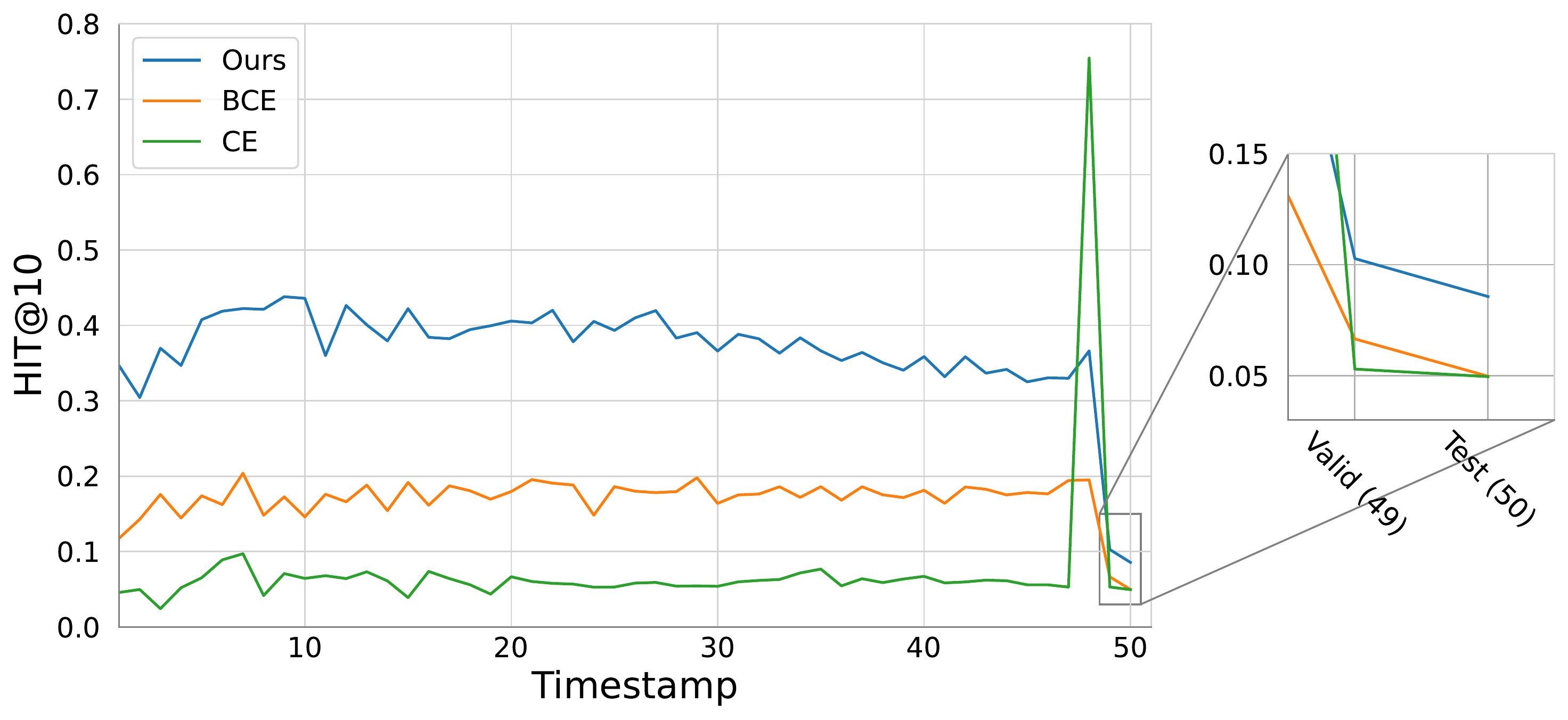}
  \caption{Evaluate the HIT@10 metric of the SASRec model on each timestamp of user sequences in the Beauty dataset, where SASRec is trained using CE, BCE, and our improved loss function, respectively. The model was trained on user sequences with a maximum length of 50, and the last two timestamps represent the validation and test set results, respectively.}
  \label{fig:why_work}
\end{figure}

\subsection{RQ2: Unpacking the Performance Boost of Improved Loss Function in Sequential Recommendation}
To address RQ2, we aim to investigate the underlying reasons  for the improved effectiveness of our approach. As discussed in Section \ref{sec:exploration}, our improved loss function extends the CE to all timestamps in the input sequence. Therefore, we compared the performance of our improved loss function, CE, and BCE at each timestamp in the sequence,
and present the detailed experimental results in Figure \ref{fig:why_work}. 
Notably, the evaluation results for timestamps 48, 49, and 50 correspond to the training set, validation set, and test set, respectively, whereas the results for the first 47 timestamps represent the model's performance at the respective timestamps.
From the figure, it is apparent that CE achieves the best result (HIT@10: 0.7546) at timestamp 48, but underperforms at all other timestamps, which confirms its optimization ability while exposing its limitations. This is because CE only optimizes the last timestamp of the training sequence. 
Conversely, the results of BCE correspond to optimizing all timestamps in the training sequence.
While it underperforms CE at timestamp 48, it slightly outperforms CE on the validation and test sets. Finally, our improved loss function significantly outperforms both BCE and CE, except for being inferior to CE at timestamp 48, but still an acceptable level. In summary, the improved performance of our approach is attributed to its robust optimization ability compared to BCE and its more comprehensive optimization compared to CE.

\section{Conclusion}
In this paper, we conducted an analysis of the advantages and disadvantages of commonly used loss functions in sequential recommendation, and proposed an improved loss function that significantly enhances the performance of sequential recommendation models. Furthermore, we established higher benchmarks for GRU4Rec, SASRec, SR-GNN, and S$^3$Rec models, with SASRec outperforming BERT4Rec on both the ML-1M and Beauty datasets. We also helped reproduce the results of BERT4Rec on the Beauty dataset. Moving forward, we plan to expand the experimental results to include more sequential recommendation models and more datasets. Ultimately, we hope that our work will inspire the research of new sequential recommendation models.



\bibliographystyle{ACM-Reference-Format}
\bibliography{paper}

\end{document}